\PassOptionsToPackage{colorlinks=true,linkcolor=blue,citecolor=blue,urlcolor=blue}{hyperref}
\documentclass[preprint,12pt]{elsarticle}

\usepackage{amssymb}
\usepackage{amsmath}
\usepackage[dvipsnames]{xcolor}
\usepackage{adjustbox}
\usepackage{float}
\usepackage{graphicx} % Required for inserting images
\usepackage{soul}
\usepackage{etoolbox} % For patching environments
\AtBeginEnvironment{tabular}{\Large}
\usepackage[utf8]{inputenc}      % for input encoding
\usepackage[T1]{fontenc}         % for font encoding
\usepackage{textcomp}    
\usepackage{hyperref}

\journal{Applied Energy}

\begin{document}

\begin{frontmatter}

\title{Investigating the emergent invariant properties of Hungarian electric distribution networks}

\author[1,3]{Michelle T. Cirunay\corref{cor1}}
\author[2]{B\'alint Hartmann}
\author[2]{T\'imea Erdei}
\author[2]{Tam\'as Soha}

\address[1]{Institute of Technical Physics and Materials Science, HUN-REN Centre for Energy Research, Budapest, Hungary}
\address[2]{Institute of Energy Security and Environmental Safety, HUN-REN Centre for Energy Research, Budapest, Hungary}
\address[3]{De La Salle University, Manila, Philippines}

\cortext[cor1]{Corresponding author: michelle.cirunay@ek.hun-ren.hu}

%% Abstract
\begin{abstract}
Electric power distribution networks serve as the final and essential stage in power delivery, bridging transmission infrastructure and end users. The structural configuration of these networks plays a critical role in determining system reliability, fault tolerance, and operational efficiency. Although the design of distribution systems is influenced by various regional factors, such as geography, customer density, and planning standards, the extent to which consistent structural characteristics emerge across different networks remains an open question. In this study, we perform a detailed spatial and topological analysis of five MV distribution networks in Hungary. Despite notable differences in geographic layout and consumer distribution, we identify statistically consistent patterns across several key metrics, including degree, BC, and powerline length. These findings suggest the influence of common underlying design principles or optimization constraints, potentially indicating universal structural tendencies in MV network design. The results provide insight into the organization of real-world distribution systems and offer a basis for improved planning, risk mitigation, and system optimization in future grid developments.
\end{abstract}

%%Research highlights
\begin{highlights}
\item Metrical and topological properties of five (5) medium-voltage (MV) distribution networks of Hungary were investigated using GIS and complex network analysis techniques.
\item These properties were related to the electric supply quality and reliability of the distribution networks.
\item It was found that the population density and the number of microvillages are more influential than population on the network efficiency and structure, respectively.
\item Despite notable differences in geographic layout and consumer distribution, statistically consistent patterns were identified across several key metrics supportive of the hypothesis of universal behavior.
\end{highlights}

%% Keywords

\end{frontmatter}

\section{Introduction}
\label{sec:intro}
If real-world infrastructure networks—such as electric grids, roads or water supply are considered as networks, their nodes and edges are embedded in the plane, therefore long connections lead to greater cost. They are generally designed based on similar optimization principles, to balance cost, efficiency, resilience, and spatial constraints ~\cite{Barthelemy2022-kn}, leading to convergent structural patterns across different types of networks.

In practice, network design usually aims to minimize the constructions cost (total network length) while maintaining as high performance (efficiency of flow) as possible. Such problems are usually solved as minimum spanning tree problems, or maximizing efficiency while fixing total network length ~\cite{Guillier2017,Bolukbasi2018,Morer2020}. These optimization models rely on very similar approaches, which is an explanation of why infrastructure network designs are homogeneous lattice-like graphs. However, while a pure minimization of total network length usually leads to tree-like networks, real-world infrastructures typically include loops to increase their robustness. This trade-off is demonstrated in several transport theory works, implying that when resilience is rewarded, loops emerge in the structure~\cite{Banavar2000,Chekuri2005,Katifori2010}.

Electric distribution is the final stage in the delivery of power, carrying electricity from the transmission system to end users. The architecture of distribution networks plays a pivotal role in ensuring efficient power flow, fault detection and isolation, reliability through redundancy, and effective load balancing~\cite{BaranIEEEPOWERDELIVERY1989}. Beyond technical functionality, the design and planning of these networks influence voltage stability, energy loss reduction, and overall system efficiency. Moreover, they contribute to economic and environmental goals by reducing operational costs and carbon emissions. In short, understanding the structural organization of distribution networks is essential for maintaining the secure, reliable, and efficient functioning of the electrical grid~\cite{AbeysingheAPPLIEDENERGY2018, HinesINTSYSSCI2010}.
Medium-voltage (MV) electric distribution networks typically use radial feeder layouts in a tree- (or star) like structure, which is a result of the above discussed cost-optimizing design approach ~\cite{Kaiser2020}. To increase resilience of these networks, optional looping is provided with tie-lines, switches, among others.

From the perspective of centrality measures, these design principles lead to a small proportion of bottleneck nodes and very few alternative routes. Primary substations behave as dominant hubs in their respective service areas, while secondary substations form less dominant ones. (It was reported in ~\cite{Louf2013} that hierarchical structures are related to hubs controlling geographically separated areas.) This leads to two BC modes. Nodes located on the MV feeders will have high BC values, and nodes located on the MV branches will have lower BC values. The presence of these two hub levels can lead to distinct centrality regimes~\cite{Louf2013}.

This bimodal BC pattern is even more accentuated by the historical spatial growth process of MV networks. A recent work by the authors shows that the expansion of MV networks resulted in peripheral nodes found at the actual peripheries, evident in the increased probability of finding nodes with very low betweenness centralities. The bimodal peaking was likely caused by the fact that powerlines were constructed from a limited number of central locations, and after major lines were built, the importance of these locations became smaller, thus also decreasing their BC~\cite{HartmannEtAlARXIV2024}. Similar results are reported in ~\cite{KirkleyNATCOMMS2018}, namely that if the network remains planar, BC distribution remains invariant even with large growth. These findings suggest that MV networks in general should behave like planar networks due to spatial embedding.

With the rise of geographic information systems (GIS), vast amounts of geo-spatial data have become available, enabling a detailed analysis of the structural and spatial characteristics of complex infrastructure systems, including electric power grids~\cite{NayeripourWORLDAPPSCI2010, AbdulrahmanSPATIALINFO2020}. When integrated with tools from network science, GIS provides the means to analyze the topology of distribution networks, revealing strong correlations between network structure and system-level risk~\cite{XuIETGENTRANS2019}. This motivates a deeper investigation into the physical and topological properties of electric distribution systems. In literature, previous works have made use of common network metrics such as electrical centrality~\cite{WangCDC2010}, topological similarity, degree distribution, clustering, diameter and assortativity, among others~\cite{AbeysingheAPPLIEDENERGY2018, HinesINTSYSSCI2010} to enhance resilience~\cite{DwivediINTJCRITICAL2024}, identify vulnerabilities~\cite{FanSYSTEMS2022, ChenIETGENTRANS2020}, and assess reliability~\cite{XuIETGENTRANS2019, WuICISCAE2024}.

\begin{figure}[ht]
\centering
\includegraphics[width=0.9\linewidth]{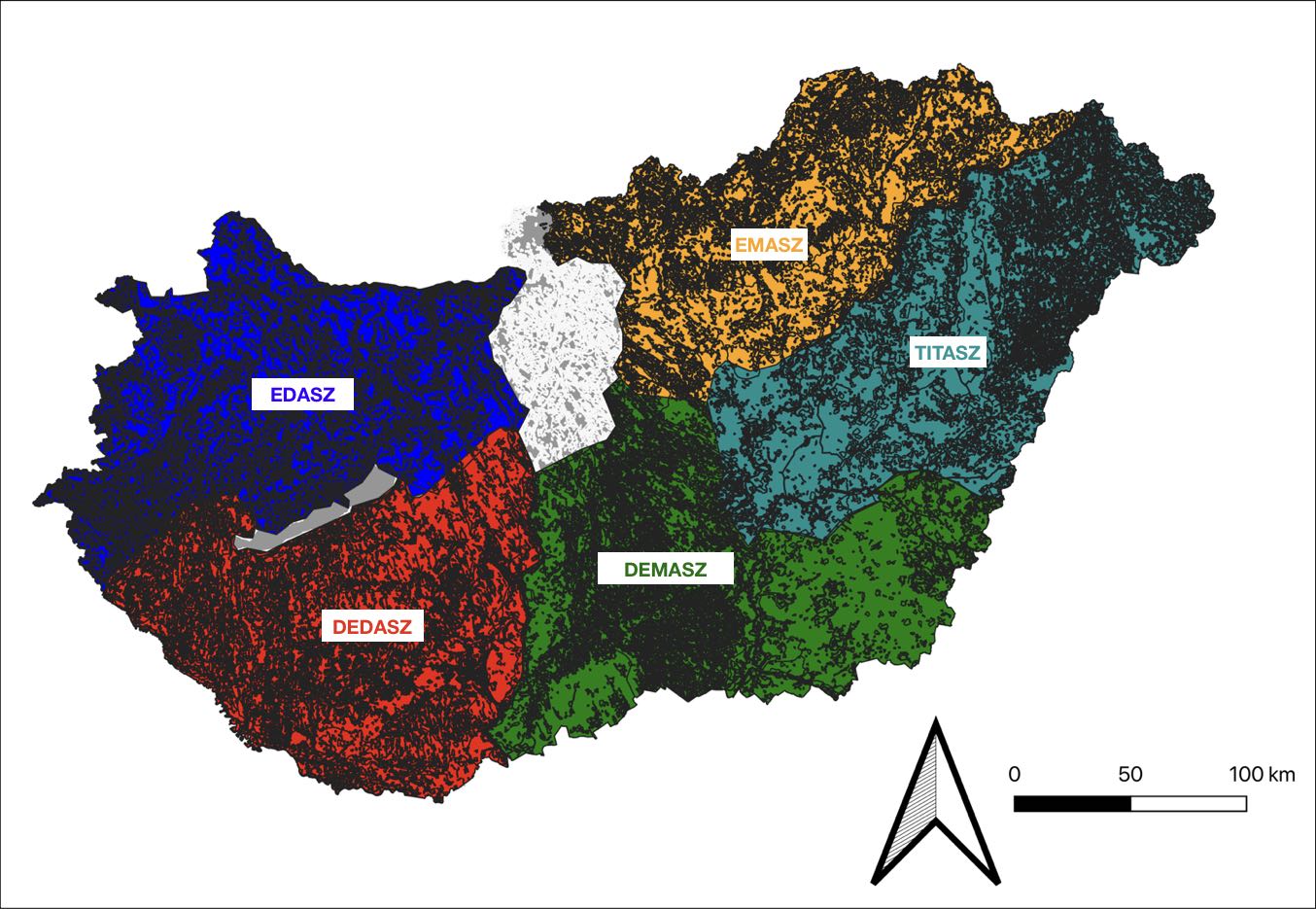}
\caption{Map of the electric distribution networks that comprise the Hungarian power grid.}
\label{fig: maps}
\end{figure}

In this study, we focus on the geophysical and network structure of five MV distribution networks managed by distribution system operators (DSOs): DEDASZ, DEMASZ, EDASZ, EMASZ, and TITASZ, which supply electricity across Hungary (see Figure~\ref{fig: maps}). Due to incomplete and unreliable data, we exclude the ELMU network serving the country's capital, Budapest (in gray).  Despite being the result of the same design process, all regional networks are shaped by a combination of technical, regulatory, geographic, and economic constraints.

Regardless of these expected variations, we find that all five networks, each serving regions of similar size and population density, exhibit remarkably consistent structural features across several topological metrics, including degree distribution, BC, and total line length. These similarities confirm the existence of common design constraints or emergent optimization principles, raising the possibility of universal patterns in MV distribution network architecture. In this work, we investigate these patterns and explore the mechanisms that may underlie their apparent invariance.

The observation of universal structural properties across diverse electric distribution networks has important implications for both the analysis and design of power systems. If such universality holds, it suggests that these networks are shaped by shared optimization principles or constraints such as minimizing cost, ensuring reliability, or adhering to spatial efficiency that transcend local geographic or regulatory differences. From an engineering perspective, this enables simplified modeling and generalization, allowing insights from one network to be applied to others with similar characteristics. Universal behavior also enhances predictive capabilities, making it possible to estimate vulnerability, performance, or expansion behavior even in regions with limited data. Moreover, common structural features may correspond to common failure modes, offering utilities a strategic advantage in designing more robust and resilient systems. This also opens the door for standardized planning practices, which could improve cost-efficiency and interoperability. On a broader level, the emergence of universal patterns invites further investigation into the fundamental processes governing the evolution of infrastructure systems, bridging insights from network science, optimization theory, and power engineering. As such, recognizing and understanding universality in distribution networks is a powerful step toward more intelligent and adaptive grid development.

The contribution of this work is three-folds: (i) The Hungarian DSO networks considered have never been investigated as they came from repaired data sources (ii) We provide a comprehensive spatial and topological characterization of these electric distribution networks; and (iii) We show that despite the differences in the details of the structure and the areas being catered by these DSO networks, similar statistical properties persist which supports the hypothesis of universal behavior.

\section{Methodology and Datasets}
\label{sec:methods}

The characterization of MV networks is critical for power system planning, reliability, optimization, and resilience as they serve as links between high-voltage (HV) transmission systems to low-voltage (LV) consumers.

As the accuracy of our description of the networks can only be as good as the data, we took some steps to ensure its quality. In the following, we describe the process of MV network repair and landcover data.

\subsection{Methods of data acquisition}

\subsubsection{Hungarian DSO Networks}

The analyses were conducted on the MV electric distribution network in Hungary. Corrections were made to the previously generated vector data to ensure that network elements appeared as continuous lines. These adjustments were necessary to facilitate accurate analysis. The process involved generating points at the end of lines, separating nearby points to eliminate errors, and establishing connection lines between these points. After automated repair using the above steps, a manual back-check was conducted to identify and correct the remaining errors.

\subsubsection{Related geographical data and preparation}

The 2018 land cover data from the Corine Land Cover database ~\cite{EUENVIAGENCY2018} was utilized for the analysis. The Corine layers were split with mesoregion boundaries, allowing for the identification of land cover patches within each ~\cite{novenyzetiterkep}. The Corine layers were categorized based on their land cover characteristics, and statistical summaries were generated for each mesoregion according to a simplified classification scheme.

A similar approach was applied to the MV network data. The grid was segmented according to mesoregion boundaries, and the network components within each region were identified. This facilitated the calculation of the total length (in kilometers) of the MV grid within each mesoregion.

The statistics on the population of mesoregions are based on the location of the centre of the polygon from settlement data, as the administrative boundaries of municipalities and mesoregional areas do not coincide. Administrative area of the municipalities covered by each DSO network was defined, and the population and area data from the latest census~\cite{kshStatinfoTheme} were associated to calculate population density and distribution. An existing characterization was applied to define the role of each settlement in the settlement hierarchy and structure in Hungary, by characterizing them as microvillages (population $ < 700$) and other settlement types~\cite{jordan2021kocsis}. For historical reasons, most of the originally equally distributed microvillages in Hungary were abandoned and later disappeared, making the population of those regions concentrated in larger settlements, mainly on the Great Hungarian Plains. It was presumed that in certain regions in Hungary, the organic development of settlements and their distribution may have affected the powerline development in the 1950s and 1960s, making it fragmented, thus having an imprint on the current layout.
Land cover, mesoregion, and settlement structure analysis were performed in ArcGIS, using basic geoprocessing operations and SQL.

\subsection{Metrical and topological parameters}

\subsubsection{Metrical parameters}
In power system terminology, the network length refers to the entire length of overhead and subsurface electrical distribution line segments~\cite{AbdulrahmanSPATIALINFO2020}. The network length is an important metric for electrical power networks because it affects a variety of technical and economic variables such as voltage drops, power losses, and the cost of cables or overhead lines in the system. It is also necessary when explaining the network topology in detail.

Forming electric distribution networks (and consequently the power grid network) is a complex process. While power grids are in some way planned to ensure their reliability, the planning and creation are subject to many factors such as the population demand, generation, and developments related to renewable energy resources, among others. 

We can view each electric distribution network as a collection of individual cable lines $\ell$ (identified via unique IDs) that connect the transmission system to the consumers. Here, each powerline can be characterized by its length $L_{\ell}$, which is the sum of the edge lengths. Each distribution network can be characterized by the powerline length $L = \sum \ell L_{\ell}$. Finally, to have a sense of the demand for each distribution network, we also account for each network's land area coverage $A_o$ and the population $P_o$ it is catering.

\subsubsection{Network Parameters}

Because of its structure, electric distribution networks can also be understood using the techniques developed in complex network theory. Graphs are mathematical entities consisting of nodes whose interactions are modeled by a connection or edge drawn between them. For this work, we create the primal representation of the networks wherein an individual powerline is a path in the network, distribution transformers, switches, busbars and consumer locations are nodes $i$, and the physical portions of the powerline between nodes $i$ and $j$ are the edges $e_{ij}$. We can therefore think of each network as a collection of nodes and edges, and measure: the number of nodes $N$ and edges $E$. 

To have a more complete analysis of the networks, we also look into other topological properties. For example, degree-related measures can tell us about the structure of the networks being considered. For example, in power networks, node degrees can be such that substations have a high node degree compared to the other nodes. Here, we explore the average node degree $\left< k \right>$, the average number of edges per node in the graph; and degree centrality distribution $P(DC)$ which is a measure of local centrality that is calculated from the immediate neighborhood links of a node (more details in the next section). An average node degree of $\left< k \right> = 2$ can signal tree-like topology, while $\left< k \right> > 2$ means the network has a meshed structure.

In addition to these, we also look at topological length-related quantities such as the average shortest path length $\left< \ell \right>$ and network diameter $D$. These quantities are derived and related to the shortest path, which is the geodesic distance between any nodes $i$ and $j$. Here, the average length along the shortest paths for every feasible pair of network nodes is determined to be the average path length $\left< \ell \right>$. This has been widely used to gauge the efficiency of information and transportation networks. On the other hand, the longest shortest path is the network diameter $D$. 

Finally, we also looked into the degree to which nodes in a graph tend to cluster together via the clustering coefficient $C$ and the strength of division of a network into modules, which is known to be the network modularity $Q$. Nodes in the majority of real-world networks (especially social networks) tend to congregate together. A radial network's clustering coefficient, however, is zero. This makes the measure useless for analyzing the MV and LV radial networks. However, this measure has been widely employed in investigations of HV networks. On the other hand, networks with high modularity have dense connections between the nodes within modules but sparse connections between nodes in different modules.

\subsubsection{Centrality Measures}

 Representing the electric distribution networks as graphs allows us to utilize the tools of complex network theory for characterizing the networks. The significance of every node in the network is usually measured using various centrality metrics, each of which has a unique interpretation depending on the particular network connectivity metric being used. For our purpose, we considered three centrality metrics, namely: (i) degree centrality $DC$, (ii) closeness centrality $CC$, and (iii) betweenness centrality $BC$.
 
 The degree centrality $DC$ can give us an idea of the structure of the network by identifying which node has the most edges incident on it. In the context of power networks, substations are nodes that have high degrees compared to the rest of the nodes in the network. 
 
 \begin{equation}
\label{eqn:degree}
\centering DC_i = \frac{1}{N-1} \sum_{j}^{N} \alpha_{ij}
\end{equation}

 where $N$ is a number of vertices and $\alpha_{ij} = 1$ if there is a direct link between nodes $i$ and $j$ where $j \neq i$.
 
 On the other hand, the other two centrality measures, closeness $CC$ and betweenness $BC$ centralities, are related to the physical layout of the nodes in space.

The closeness centrality $CC$ is given by

\begin{equation}
\label{eqn:closeness}
\centering CC_i = \left[ \sum d_{ij}\right]^{-1} 
\end{equation}

For all $j \neq i$; this metric gives the structural center of the network, as the node with the highest closeness centrality has the minimum shortest path length, $d_{ij}$, to all the other nodes. On the other hand, the BC is given by

\begin{equation}
\label{eqn:betweenness}
\centering BC_i = \sum \left[\frac{\sigma_{hj}(i)}{\sigma_{hj}} \right]
\end{equation}

for all $h \neq j \neq i$. It measures how many times the node $i$ has appeared,$\sigma_{hj}(i)$, in the collection of all shortest paths in the network,$\sigma_{hj}$. The high-$BC$ nodes, therefore, may be thought of as bridges where efficient connections between peripheral nodes pass through.

\section{Properties of the Distribution Networks}
\label{sec:results}

\subsection*{Metrical and topological characterization}

\begin{table}[]
\centering
\begin{adjustbox}{width=\textwidth,center=\textwidth}
\setlength{\tabcolsep}{10pt} % Default value: 6pt
\renewcommand{\arraystretch}{2} % Default value: 1
\begin{tabular}{|c|c|c|c|c|c|l|l|l|}
\hline
\textbf{Networks} &
  \textbf{\begin{tabular}[c]{@{}c@{}}Total Population\\ $P_o$\end{tabular}} &
  \textbf{\begin{tabular}[c]{@{}c@{}}Total Area $A_o$\\ {[}km$^2${]}\end{tabular}} &
  \textbf{\begin{tabular}[c]{@{}c@{}}Population\\ Density\\ $P_o/A_o$ \\ {[}1/km$^2${]}\end{tabular}} &
  \textbf{\begin{tabular}[c]{@{}c@{}}MV\\ Line Length\\ $L_{total}$\\ {[}km{]}\end{tabular}} &
  \textbf{\begin{tabular}[c]{@{}c@{}}Line \\ Density\\ $\lambda$\\ {[}1/km{]}\end{tabular}} &
  \textbf{\begin{tabular}[c]{@{}l@{}}Number of\\ Settlements\end{tabular}} &
  \textbf{\begin{tabular}[c]{@{}l@{}}Number of\\ Microvillages\\ ( \textless 700 people)\end{tabular}} &
  \textbf{\begin{tabular}[c]{@{}l@{}}Share of \\ Microvillages \\ {[}\%{]}\end{tabular}} \\ \hline
\textbf{DEDASZ} & 1,196,548 & 18,333 & 65.3  & 10,323 & 0.56 & 842 & 763 & 91 \\ \hline
\textbf{DEMASZ} & 1,327,017 & 18,213 & 72.9  & 11,786 & 0.65 & 262 & 0   & 0  \\ \hline
\textbf{EDASZ}  & 1,754,618 & 18,226 & 96.3  & 12,731 & 0.70 & 868 & 628 & 72 \\ \hline
\textbf{EMASZ}  & 1,321,902 & 15,437 & 85.6  & 8,415  & 0.55 & 650 & 416 & 64 \\ \hline
\textbf{TITASZ} & 1,481,803 & 18,718 & 79.2  & 11,044 & 0.59 & 395 & 120 & 30 \\ \hline
\end{tabular}
\end{adjustbox}
\caption{\label{tab:metricalprops}Metrical properties and settlement data of the DSO networks}
\end{table}

The structure of electric distribution networks is closely linked to the demographic characteristics and land cover of the regions they serve. Consequently, they can hint at the performance of the networks in terms of electric supply reliability. To gain some insights into the relationship between demographics, infrastructure needs, current network state, and supply reliability we present a comparative overview of the metrical, topological properties, and electric supply quality of the regions and the Hungarian DSO networks in Tables~\ref{tab:metricalprops},\ref{tab: networkprops}, and \ref{tab:supply-quality}.

Since we are dealing with MV distribution networks, notice that in Table~\ref{tab: networkprops}, all the networks have an average degree $\left< k \right> \approx 2$ and clustering coefficients $C \sim 0$, which are characteristics of radial or tree-like topology. This topology is the simplest and cheapest for an electrical grid. However, if a line is disconnected, all the lines downstream will also lose power~\cite{PrakashAPWC2016}. Previous works~\cite{AbeysingheAPPLIEDENERGY2018} have mentioned that the clustering coefficient is useless when studying MV and LV networks. However, since there is still an observable difference in the magnitude of this metric among the DSO networks in Hungary (as shown in Table~\ref{tab: networkprops}), we still find value in including it for comparison between the structure and performance.

EDASZ emerges as the most densely populated, with a population density of 96.3 persons/km$^2$ and the highest total population at approximately 1.75M. Expectedly, as there is a bigger demand in this region, the EDASZ network has the highest MV line density at 0.70 1/km$^2$, suggesting an intensive infrastructure network tailored to a large and concentrated population. Given this, one may think that the EDASZ region is urban. However, we find that EDASZ caters to the largest number of settlements, 72\% of which are microvillages (areas with population $< 700$ people). To cater to the dense population, the EDASZ network has the highest MV line length and density. Topologically, it displays a balance between size and connectivity with a moderate average path length (and consequently, its diameter), but relatively high average degree and modularity. Such properties suggest a balanced design for a resilient power distribution system. It combines fault tolerance, decentralized control, and manageable complexity, which are all essential for handling both localized faults and large-scale disturbances. However,  Table~\ref{tab:supply-quality} shows that it exhibits relatively weak performance across most electric supply reliability measures. It is shown to have high fault/length (7.18 disturbances/100 km), higher frequency, and duration of outages. While its restoration rate is good (92.47\%),due to the significant number of microvillages to cater to, it still has considerably longer restoration time. This may be because a high modularity could mean coordination issues, Limited cross-module support, and slower global optimization.

On the other hand, DEDASZ has the lowest population density at 65.27 persons/km$^2$ with the lowest total population (1.19M) and significant land area. Despite its lower population, it has a high number of settlements (842), and an overwhelming 91\% of these are microvillages. This suggests a highly fragmented and dispersed rural settlement pattern, with many small communities likely requiring localized utility solutions. Its MV line density is relatively low (0.56 1/km$^2$), which may be due to shorter distances (low $\left< \ell \right>$) between closely clustered microvillages (high-$C$ and high-$\left< k \right>$). Despite the large land area, the high clustering of nodes results in a shorter average path length (and network diameter) and high modularity. These indicate efficient flow and easier fault isolation. As a result, the DEDASZ network has a moderate level of reliability in terms of electric supply quality with a relatively low outage frequency (0.6) and fault rate (5.78), paired with a good restoration rate (91.86\%) (shown in Table~\ref{tab:supply-quality}). However, due to the large land area of DEDASZ, it may be difficult to reach further sections, resulting in the longest restoration time among all the networks being considered.

DEMASZ is quite distinct from the others, with the least number of settlements (262) and no microvillages, indicating a much more centralized region. In this region, population concentrations arose based on agricultural production, in which urban functions were already present, without being able to rely on a significant hinterland. However, the spaces between the sparsely located market towns with large borders were filled by a specific type of scattered settlement, the farm world, which became increasingly dense in the 19th century. It has a relatively high population (1.33M) and a moderate population density (72.86 persons/km$^2$). The absence of microvillages and the highest MV line length per settlement suggest a more consolidated infrastructure network designed for larger, more populous communities. DEMASZ has a moderate performance when it comes to electric supply quality. Although it only has a tolerable level of outage frequency (compared to EDASZ and TITASZ), the restoration rate (87.01\%) is the worst among all the networks. This can be attributed to several factors, such as the network size (large network diameter and long average path length) as the longest average MV feeder length is in this region. Additionally, DEMASZ has a high clustering and low modularity, making it difficult to isolate faults and service some areas. 

TITASZ exhibits a mix of rural and urban properties with a fair population density and number of microvillages. This balance is reflected in its infrastructure with an adequate line density of 0.59/km$^2$, which suggests a mixed strategy of central service hubs and peripheral rural supply. In terms of its electric supply quality, TITASZ is shown to display mixed performance: It has the highest restoration rate after EMASZ (93.97\%) and the shortest restoration time (2.00), but it is compensated by frequent (0.93) and long outages (52.3), along with the highest number of faults per unit length (7.53). This suggests that although TITASZ restores power quickly, it suffers from frequent and prolonged outages, likely due to more fault-prone infrastructure.

Finally, EMASZ also reflects a semi-rural profile. Though it has fewer people than DEMASZ and TITASZ, it has a relatively high population density (85.63 persons/km$^2$), 650 settlements, and a significant portion of microvillages. Interestingly, it has the lowest MV line density (0.55 1/km$^2$) and line length, which might suggest challenges in network coverage due to geography ( e.g. this region covers the North Hungarian Mountains, the most mountainous part of the country)  or planning inefficiencies, despite having a considerable number of small settlements. In terms of network properties, EMASZ has the largest network diameter (and consequently, the largest average path length), the least average degree, clustering, and modularity. These characteristics reflect a stretched, radial network structure typical of rural or semi-rural areas. In such regions, the need to reach many small settlements results in sparse connectivity and increased infrastructure demands. Additionally, a low network modularity makes fault isolation difficult. However, as we can observe in Table~\ref{tab:supply-quality}, EMASZ seems to be the most reliable electric distribution network in Hungary with the lowest frequency of outages, shortest outage durations, and highest restoration rate. This may be attributed to the fact that EMASZ has the shortes MV line length and lowest density, which may be the reason for the ease of repair.

Overall, EMASZ appears to be the most robust and efficient, combining reliability with quick restoration. TITASZ, despite fast restoration, struggles with system reliability due to frequent faults and longer outages. Finally, EDASZ  exhibits the worst performance across most of the electric supply quality metrics.

\begin{table}[H]
\begin{adjustbox}{width=\columnwidth,center}
\centering
\setlength{\tabcolsep}{10pt} % Default value: 6pt
\renewcommand{\arraystretch}{2} % Default value: 1
\begin{tabular}{|c|c|c|c|c|c|c|c|}
\hline
\textbf{Networks} &
  \textbf{\begin{tabular}[c]{@{}c@{}}Number of nodes\\ $N$\end{tabular}} &
  \textbf{\begin{tabular}[c]{@{}c@{}}Number of edges\\ $E$\end{tabular}} &
  \textbf{\begin{tabular}[c]{@{}c@{}}Average node degree\\ $\left<k \right>$\end{tabular}} &
  \textbf{\begin{tabular}[c]{@{}c@{}}Average shortest path length\\ $\left<\ell \right>$\end{tabular}} &
  \textbf{\begin{tabular}[c]{@{}c@{}}Clustering coefficient\\ $C$\end{tabular}} &
  \textbf{\begin{tabular}[c]{@{}c@{}}Network diameter \\ $D$\end{tabular}} &
  \textbf{\begin{tabular}[c]{@{}c@{}}Network modularity\\ $Q$\end{tabular}} \\ \hline
\textbf{DEDASZ} & 53,185  & 55,176  & 2.0749 & 255.4831  & 0.00214 & 747  & 0.6064 \\ \hline
\textbf{DEMASZ} & 146,108 & 146,959 & 2.0116 & 673.6870  & 0.00101 & 1,958 & 0.5698 \\ \hline
\textbf{EDASZ}  & 89,360  & 90,230  & 2.0195 & 394.9072  & 0.00091 & 1,174 & 0.6051 \\ \hline
\textbf{EMASZ}  & 101,655 & 101,900 & 2.0048 & 1074.9772 & 0.00002 & 3,346 & 0.5592 \\ \hline
\textbf{TITASZ} & 69,825  & 70,395  & 2.0163 & 376.5683  & 0.00025 & 1,227 & 0.5892 \\ \hline
\end{tabular}
\end{adjustbox}
\caption{\label{tab: networkprops} Network properties of the DSO networks}
\end{table}

\begin{table}[H]
\begin{adjustbox}{width=\columnwidth,center}
\centering
\setlength{\tabcolsep}{12pt} % Default value: 6pt
\renewcommand{\arraystretch}{2} % Default value: 1
\begin{tabular}{|c|c|c|c|c|c|}
\hline
\textbf{Networks} &
  \textbf{\begin{tabular}[c]{@{}c@{}}Ave. frequency\\ of unplanned outages\\ (outages/consumer/yr) \end{tabular}} &
  \textbf{\begin{tabular}[c]{@{}c@{}}Ave. duration\\ of unplanned outages\\ (hours/consumer/yr)\end{tabular}} &
  \textbf{\begin{tabular}[c]{@{}c@{}}Rate of restoration\\ customer in case \\ of outages longer \\ than 3 minutes (\% )\end{tabular}} &
  \textbf{\begin{tabular}[c]{@{}c@{}}Faults per \\ network length\\  (disturbances/100km) \end{tabular}} &
  \textbf{\begin{tabular}[c]{@{}c@{}}Restoration time\\ for network failures\\ (hours/failure)\end{tabular}} \\ \hline
\textbf{DEDASZ} & 0.60 & 41.16 & 91.86 & 5.78 & 2.57 \\ \hline
\textbf{DEMASZ} & 0.65 & 38.12 & 87.01 & 7.07 & 2.26 \\ \hline
\textbf{EDASZ}  & 0.67 & 44.01 & 92.47 & 7.18 & 2.46 \\ \hline
\textbf{EMASZ}  & 0.58 & 33.63 & 94.56 & 5.12 & 2.01 \\ \hline
\textbf{TITASZ} & 0.93 & 52.30 & 93.77 & 7.53 & 2.00 \\ \hline
\end{tabular}
\end{adjustbox}
\caption{Electric supply quality of the Hungarian DSO Networks~\cite{MEKHEVAL2023} (averaged for the years 2021-2023).}
\label{tab:supply-quality}
\end{table}

The structure of electric distribution networks is closely linked to the demographic characteristics and land cover of the regions they serve.While this would suggest that the examined five service regions behave differently, similarities are striking, and this finding is partially contradictory to relevant literature.

In recent years, several algorithmic planning models have been published, which use land cover or land use data as input for creating optimal MV distribution networks. In their paper, Xu et al.~\cite{Xu2023} classify the rasterized service area into different land-use categories, assigning a construction cost per raster cell, depending on the actual land use. They use this grid cost as an objective function for a heuristic planning model. A similar approach is shown by Rüde et al.~\cite{Rde2024}, who combine almost 8,000 real-world MV feeders with land cover data in a regression model. They also confirm that land cover significantly impacts the capital investment needs of MV networks. The integration of GIS data is also demonstrated in some studies ~\cite{Mehrtash2020, Techane2022, Ameling2024, 10863461}. In summary, there are advanced tools for planning distribution networks in consideration of land cover. However, they rely on a 'snapshot', while real-world distribution infrastructure has been evolving for a century~\cite{HartmannHDL2024}. This suggests that the briefly presented methods may underperform in capturing the topological and statistical properties.

\begin{figure}[H]
\centering
\includegraphics[width = 0.9\linewidth]{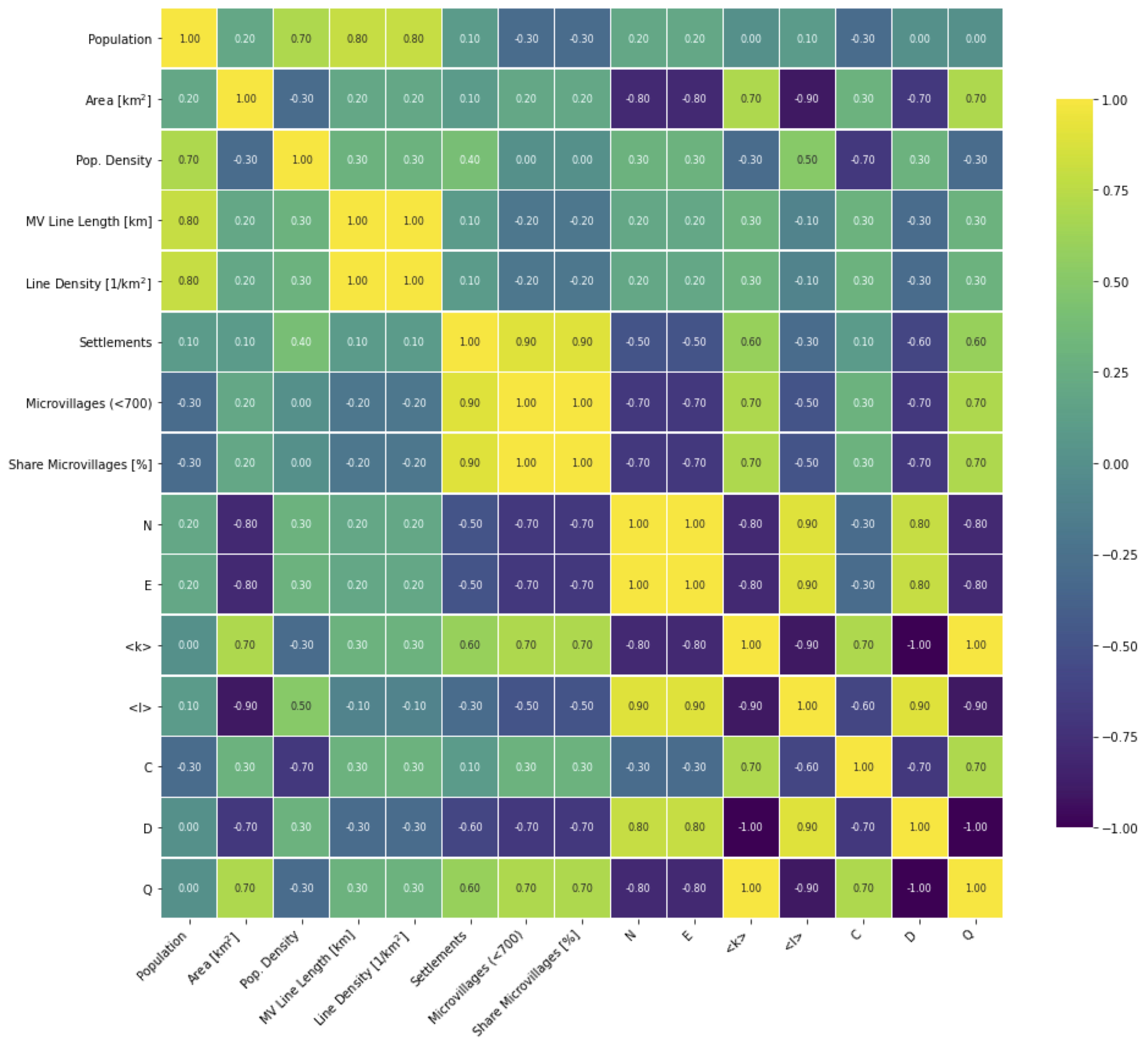}
\caption{\textit{Spearman Correlation Matrix of the DSO metrical and topological properties.}}
\label{fig:correl_metrical_topological}
\end{figure}

In Figure~\ref{fig:correl_metrical_topological} we present a Spearman correlation matrix of the metrical (demographics, landcover) and topological properties of the DSO networks. Here, we choose to use the Spearman coefficient over the Pearson coefficient because it measures the monotonic relationship between two variables where the relationship is not strictly linear, and doesn't require normality assumptions. Additionally, it is more robust to outliers because it works with ranks rather than actual values, making it suitable for noisy or skewed data. The correlation analysis reveals strong relationships between demographic characteristics, land cover, and the topological structure of electric distribution networks.

One of the most significant findings is the influence of population density on network efficiency. One may expect that a higher population density would mean shorter average path lengths, higher clustering coefficients, and greater modularity. This suggests that compact, densely populated areas support more interconnected and resilient distribution networks, which are more efficient and better equipped to handle faults. However, counterintuitively, notice that in Figure~\ref{fig:correl_metrical_topological}, we see that for the case of our datasets, there is only a moderate correlation between the population density and average path length; a strong negative correlation to clustering coefficient $C$; and a moderate negative correlation to modularity $Q$.

Additionally, Figure~\ref{fig:correl_metrical_topological} shows that the number of microvillages is more influential on the networks structure than total population, as indicated by the magnitude (regardless of sign) of the correlations, indicating that how people are distributed across a region matters more than how many people there are. For example, the number and percentage of microvillages have stronger (negative or positive) correlations to average length, network diameter, and clustering coefficients. In contrast, the population has very weak to no correlation with these network metrics. Such a result may be counterintuitive, as node degree distributions of HV networks tend to correlate with the population. However, note that we are dealing with MV grids.

While correlations are not causations, overall, these correlations highlight the importance of aligning grid design with regional demographic and geographic characteristics. Urban and dense regions benefit from meshed, modular networks, while rural areas with many microvillages may require alternative solutions like decentralized energy systems. Strategic investment in infrastructure, particularly in improving line density, can also enhance resilience and service quality in more sparsely populated or geographically challenging areas.

In the remainder of the text, we explore the probability distributions of the metrical and topological properties of the DSO networks, and we show that although these networks and the regions they cater to are very different from each other, they exhibit similarities across their statistical properties, suggestive of universality.

\subsection{Invariant Statistical Properties}

\begin{figure}[ht]
\centering
\includegraphics[width = \linewidth]{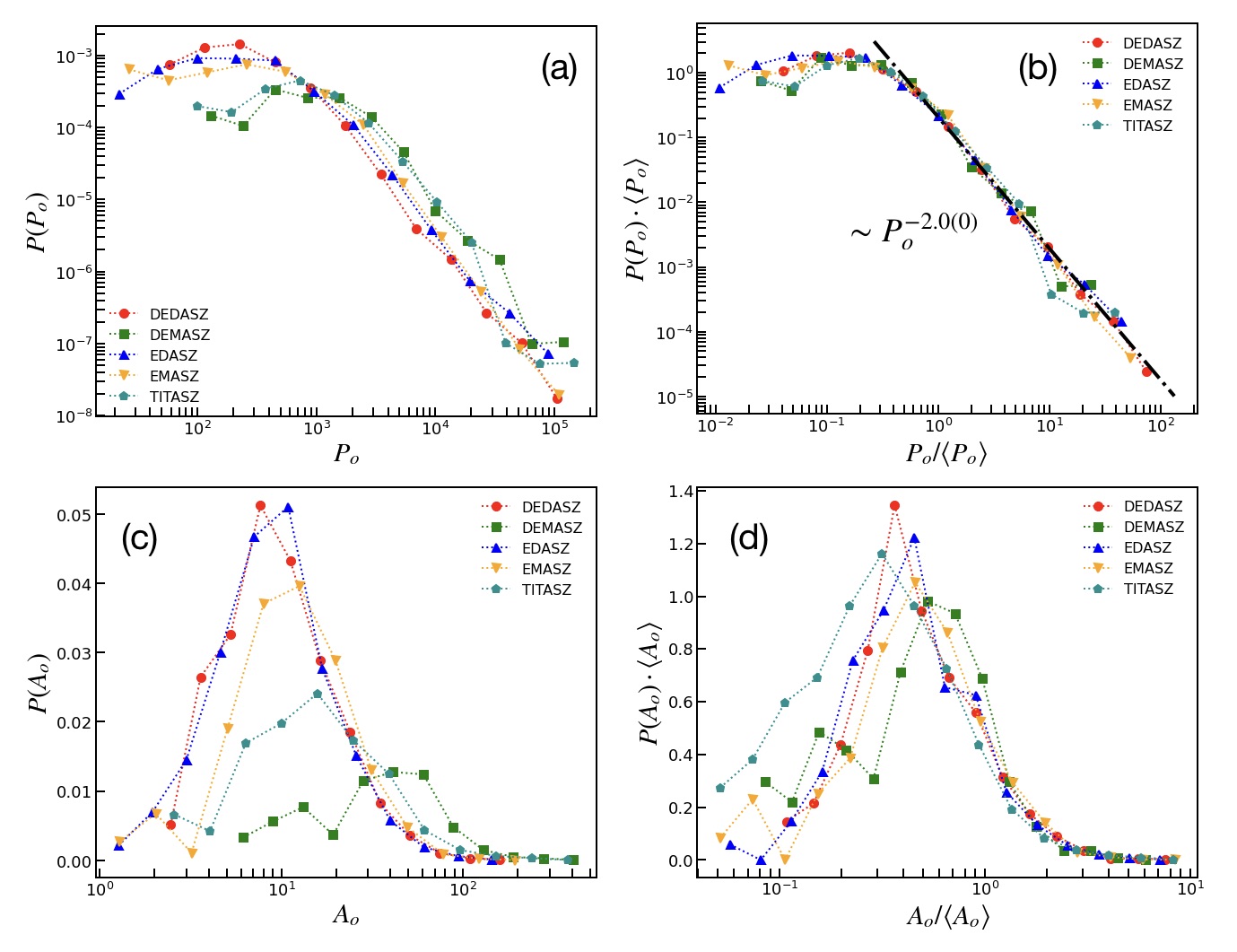}
\caption{\textit{Population and area distributions of the regions being catered by the DSO networks.} The population distributions [top panels] $P_o$ follow a PL behavior that scales according to $\sim P_o ^{-2.0(0)}$. On the other hand, the landareas [bottom panels] are unimodally distributed, indicative that there is a characteristic size in how space is divided in each region.}
\label{fig:population_area}
\end{figure}

Figure~\ref{fig:population_area} shows the probability distributions of the population and the land area across the five regions being catered by the DSO networks. Here, we find that the population (Figure~\ref{fig:population_area}(a)-(b)) across the areas scale according to a powerlaw with exponent of $\sim -2.0(0)$ while the areas of settlements (Figure~\ref{fig:population_area}(c)-(d)) follow a lognormal behavior. In literature, the probability distribution of urban areas follows a power-law behavior which scales to an exponent of $\sim -1.9$~\cite{CirunayCOMPSYS2021, CirunayIJMPC2018, LammerPHYSICAA2006, BarthelemyPRL2008}. However, note that here, this area distribution pertains to both urban and rural land areas, which explains why we are observing a lognormal behavior instead.

When the population distribution follows a power-law and the area distribution follows a lognormal distribution, it implies distinct patterns in the spatial and demographic characteristics of a system. A power-law population distribution means that a small number of regions or settlements have very high populations, while most areas have relatively low populations. This results in a high level of inequality in population sizes, with a few large urban centers dominating the overall population and many smaller settlements scattered throughout. On the other hand, a lognormal distribution for area suggests that most settlements have moderate-sized areas, with a few settlements having extremely small or large areas. This indicates that while most areas are relatively homogeneous in size, there is a long tail toward larger settlements.

The combination of these two distributions suggests a mismatch between population and area. Large settlements with high populations may not necessarily have large areas, and smaller regions may have very dense populations. This could indicate that urban centers are densely populated but may not be as spatially large, while rural areas, though expansive, may have fewer people. From an infrastructure and energy distribution perspective, this dynamic creates challenges: urban areas with high populations might require more concentrated and high-capacity infrastructure, while larger but sparsely populated rural areas would need more dispersed solutions.

Additionally, this combination highlights inequality in both population and area. While the population is concentrated in a few large centers, the distribution of area is more even, meaning that vast rural regions, though large, are underpopulated. This discrepancy can result in inequitable resource distribution, where urban areas with high populations receive more attention and services, while rural regions with larger areas may be underserved despite their greater land size. Overall, this combination of distributions suggests a highly urbanized core surrounded by expansive rural regions, each with unique infrastructure and service needs based on their demographic and spatial characteristics.

\begin{figure}[ht]
\centering
\includegraphics[width=0.9\linewidth]{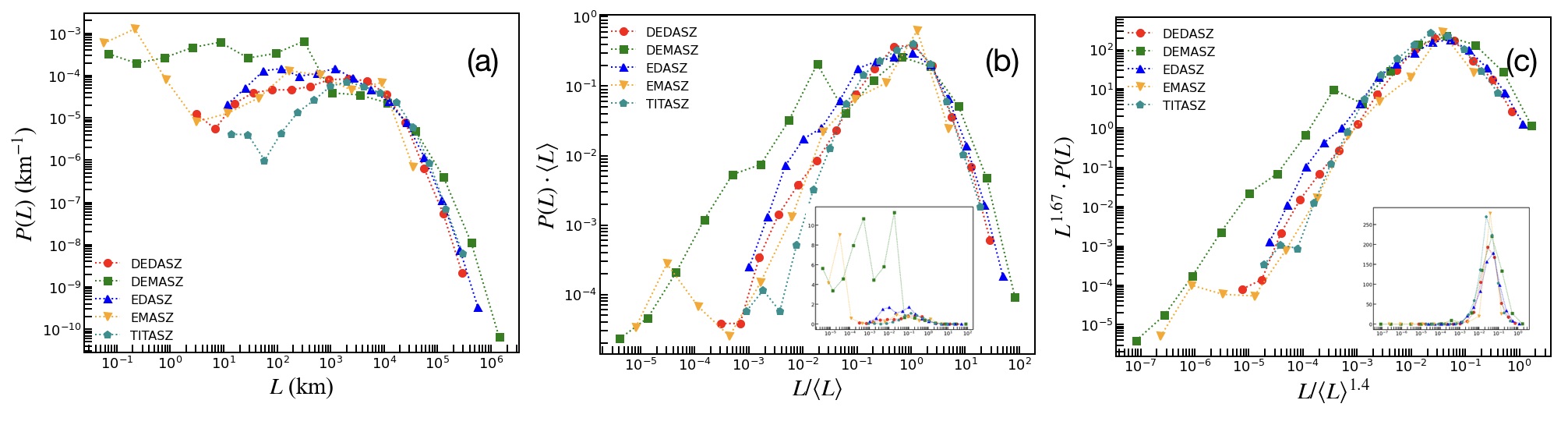}
\caption{\textit{Powerline length distributions of the DSO networks.}(a) The length distributions $P(L)$ of the electric distribution networks mostly follow unimodal trends, suggesting similar mechanisms of creation. (b) Rescaling the distributions by their respective mean did not cause the distributions to collapse. However, doing so further highlights the deviations of EMASZ in the short-length regime and the abundance of intermediate-length lines for DEMASZ. (c) Obtaining the universal scaling function by plotting $L^{\gamma} \cdot P(L)$ by $L/\left<L \right>^{\alpha}$, where $\gamma = 1/(2 - \alpha)$ (here, $\alpha = 1.4$) shows the mode and tails of the distributions collapse which hints that all the electric distribution networks of Hungary despite differences in their degree of urbanization and demands share a universal distribution after rescaling. \textit{Inset:} In a semilog scale, the lognormal behavior of the powerline length distributions becomes more apparent.}
\label{fig: lengths}
\end{figure}

Since we know the inequality introduced by the distribution of populations across land areas, it is also interesting to see how the lengths of the powerlines are distributed. Figure~\ref{fig: lengths} shows that the powerline lengths follow unimodal fat-tailed distributions $P(L)$. The unimodality of the distributions suggests a similar mechanism for the link creation even in the case of the absence of a strong, centralized planning and strict zoning rules. 

To remove the variation across our datasets, we tried different normalization approaches to collapse our data shown in Figure~\ref{fig: lengths}(b) via the respective mean of each distribution; and (c) universal scaling function utilized first by Strano et. al. on road networks~\cite{StranoRSOCOPENSCI2017}. Data collapse is a technique for determining scaling and obtaining related exponents in problems that exhibit self-similar or self-affine features, such as in critical phases, equilibrium or non-equilibrium phase transitions, complex system dynamics, and many more~\cite{BhattacharjeeJPHYSA2001}. It implies that the universal behavior is independent of system size and thus they belong to the same universality classes.

As shown in Figure~\ref{fig: lengths}(b), the distributions did not collapse when we rescaled each by its respective mean values. In Figure~\ref{fig: lengths}(c), we employ the universal scaling function $F(\cdot)$ which can be obtained by plotting $L^{\gamma} \cdot P(L)$ with $L/\left< L \right> ^{\alpha}$, where $\gamma = 1/(2- \alpha)$ ensures normalization. In literature, the universal scaling function $F(\cdot)$ has been first utilized on the length distributions of the Global Road Network (GRN)~\cite{StranoRSOCOPENSCI2017}. Their results provide evidence for general statistical laws describing global road lengths conditional on land use. At the global scale, urban and cropland roads share a universal distribution after rescaling that satisfies $\alpha = \gamma = 1$. We believe that the same scaling ansatz can be applied for powerline lengths since road networks and power distribution networks are closely related because, for easier access, maintenance, and planning~\cite{ArderneSCIENTIFICDATA2020, SadhuENERGYINFO2022}.  As previously mentioned, they are generally designed based on similar optimization principles, to balance cost, efficiency,
resilience, and spatial constraints, leading to convergent structural patterns across different types of networks~\cite{Barthelemy2022-kn}. For example, roads provide practical routes for laying overhead lines or underground cables and are used as shared corridors for utilities. Additionally, it makes sense for coordination to happen when either infrastructure is under development to ensure efficient design and regulatory compliance, especially in both urban and rural settings.

In Figure~\ref{fig: lengths}(b), we attempt the same rescaling  ($\alpha = \gamma = 1$,) i.e., rescaling by the simple mean of powerline length $L/\left< L \right>$  of the data, which is observed to not perfectly collapse the data. This shows that although road networks and power grid networks (and the constituent electric distribution networks) are planar graphs embedded in space and are created to service distant locations, the creation of powerlines is different from how roads are built in space. For one, the cable lengths are proportional to power losses and voltage drops. The presence of characteristic lengths represented as the peak tells us that there seems to be an optimal length that can minimize these quantities~\cite{1601448,Sun1982,1033739,1295010,1583718,1717599}.

In Figure~\ref{fig: lengths}(c), we utilized the case of $\alpha = 1.4$ (and, thus, $ \gamma \sim 1.67$), which resulted in all the distributions collapsing under a single curve. This time, we can still observe a scaling law but with non-trivial exponents. For example, $\alpha=1.4$ (scaling for $L$)  which implies that as the mean powerline length increases across regions, the actual segment lengths scale superlinearly indicating higher average powerline lengths (e.g., rural or sparse grids) and more long lines which may be due to geographic sparseness (e.g., rural areas), centralized power generation supplying distant areas, and planning strategies that prioritize long, HV transmission lines in certain areas. On the other hand, the tail exponent $\gamma = 1.67$ indicates that the distribution of powerline segment lengths is heavy-tailed. This means that while most segments are relatively short, such as local distribution lines, there is also a non-negligible number of very long transmission lines, like regional or national HV lines. This heavy-tailed distribution is a characteristic feature of hierarchical infrastructure systems, where local grids are densely meshed with shorter powerlines, while longer transmission lines span large distances to connect more sparsely located areas.

The fact that the powerline distributions of the five DSO networks collapse under the same curve suggests that universal principles govern the structural properties of these networks: (i) hierarchical structure: This means that, across different regions, power distribution networks tend to have a dense core of short, local lines (e.g., serving neighborhoods) and fewer long transmission lines that cover larger distances (e.g., connecting cities or regions) (ii) similar design principles: Despite regional differences, the networks likely follow common engineering goals—efficiently connecting power sources to consumers while balancing cost, land use, and the need for long-distance transmission. (iii) common scalability: The consistent heavy-tailed distribution ($b=1.67$) shows that powerline networks grow predictably with high probability of short lines and few long ones, regardless of region, as demand or area increases.

As a result, we observe a noticeable deviation of the DEMASZ network in the short-to-mid length region. From Table~\ref{tab:metricalprops}, we have an idea that DEMASZ is much more centralized, as there are no microvillages and a higher number of MV line lengths suggestive of a more consolidated infrastructure network designed for larger, more populous communities.

By collapsing the distributions, the invariance in the behavior of the powerline length distributions is more apparent, as we can see that they all follow a lognormal behavior. This indicates that most of the powerlines are of moderate length, with a smaller number of very short or very long lines. This type of distribution suggests a skewed infrastructure, where the majority of the network consists of standard-length lines, but there are some sections of the grid with significantly longer powerlines, likely due to geographic challenges or the need to service remote areas. The implications of this include higher costs and more maintenance requirements for the longer lines, as they are more exposed to environmental factors and may be harder to reach for repairs. These longer powerlines, often located in remote or sparsely populated areas, can reduce the overall efficiency of the network due to voltage drops, energy loss over distance, and potential capacity constraints. Furthermore, the resilience of the network could be impacted, as faults in these long lines may result in larger, more widespread outages, with longer detection and repair times. This lognormal distribution also reflects the structure of the existing grid, where most powerlines are already established, but longer lines extend to less-developed or emerging areas. For future grid expansion, this suggests that while new powerlines may mostly be shorter connections, significant attention will be needed to address the challenges posed by long powerlines to ensure efficient, reliable service in more difficult-to-reach areas. Overall, a lognormal distribution of powerline lengths highlights both the efficiency of the central network and the challenges posed by extended lines that require extra attention in terms of maintenance, resilience, and cost.

\begin{figure}[ht]
\centering
\includegraphics[width=0.9\linewidth]{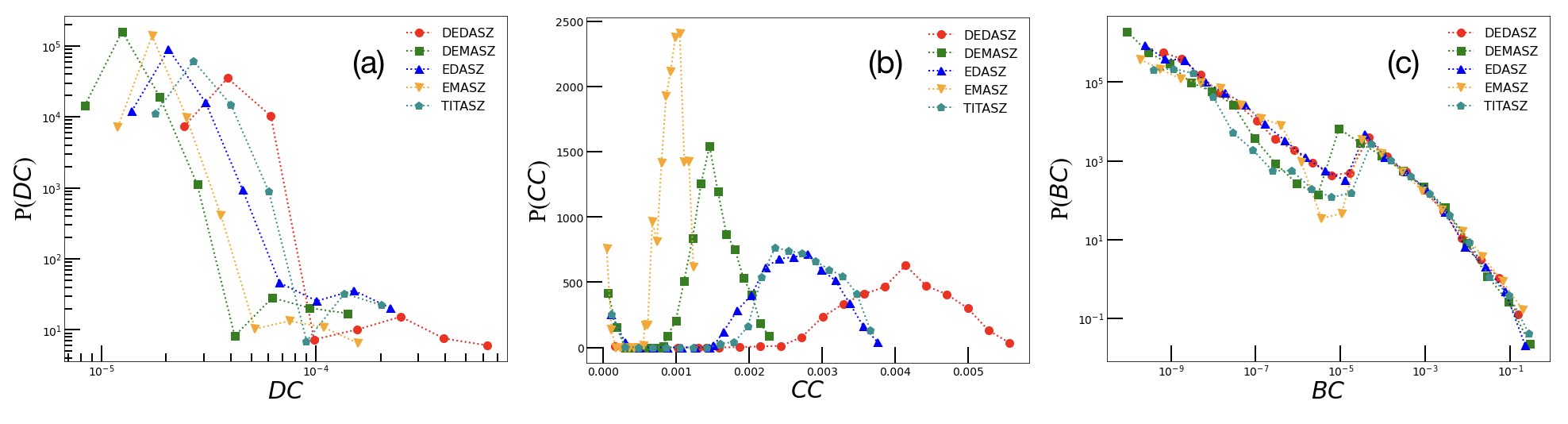}
\caption{\textit{Statistical distributions of centrality measures:} (a) degree centrality ($DC$) (b) closeness centrality ($CC$) (c) betweenness centrality ($BC$).}
\label{fig: centralities}
\end{figure} 

We also looked at the distributions of the degree $DC$, closeness $CC$, betweenness centralities $BC$ of the networks, Figure~\ref{fig: centralities}. As previously mentioned, the degree $DC$ centrality can give us an idea of the network structure. On the other hand, the closeness $CC$ and betweenness $BC$ values are the most associated with physical space among all centrality measures. 

Figure~\ref{fig: centralities}(a) shows the DC distributions follow lognormal behavior, indicating that most nodes in the network have a degree close to a central value, with a few nodes having either very few or many connections. This results in a network where connectivity is generally moderate, and there is a mix of well-connected and sparsely connected nodes. Additionally, we observe a progression of the peaks of the distributions towards high-DC values that may be correlated to the level of urbanization of the region that the electric distribution network supplies. For example, the DEDASZ has high-probability of nodes with high-DC consistent with the fact that it has the maximum average degree $\left<k \right>$, clustering coefficient $C$, and modularity $Q$ described in Table~\ref{tab: networkprops} hinting at a more compact network.

Figure~\ref{fig: centralities}(b) shows the CC distributions. The closeness metric gives us the structural center of the network. By its definition as the inverse of the shortest path length, nodes with high degree of closeness centrality can reach all other nodes in the network with ease. Intuitively, the larger the spatial span of the network, the $P(CC)$ distributions are expected to shift to smaller values. Take, for example, the case of EMASZ and DEDASZ, which are the largest and smallest networks (based on network diameter $D$) such that EMASZ has its closeness centrality distribution peaking at the lower values, and DEDASZ peaks at the larger values. Here, we find that the closeness centrality distribution is normally distributed, which implies that most nodes have similar efficiency in terms of accessing the rest of the network, meaning that the network is relatively efficient and most nodes are similarly positioned in terms of proximity to others.

Finally, the BC distributions of the DSO networks are shown in Figure~\ref{fig: centralities}(c) follow a truncated power-law consistent with the expected behavior for random planar graphs~\cite{KirkleyNATCOMMS2018}, which suggests a network with a moderately centralized and balanced structure. However, the truncated power-law in BC highlights a key aspect of the network: a small number of nodes play a critical role in connecting different parts of the network, but their centrality is capped. This capped centrality reduces the risk of extreme bottlenecks, but it still suggests that removing these high-BC nodes could disrupt communication across the network to some extent.

\begin{figure}[H]
\centering
\includegraphics[width=0.9\linewidth]{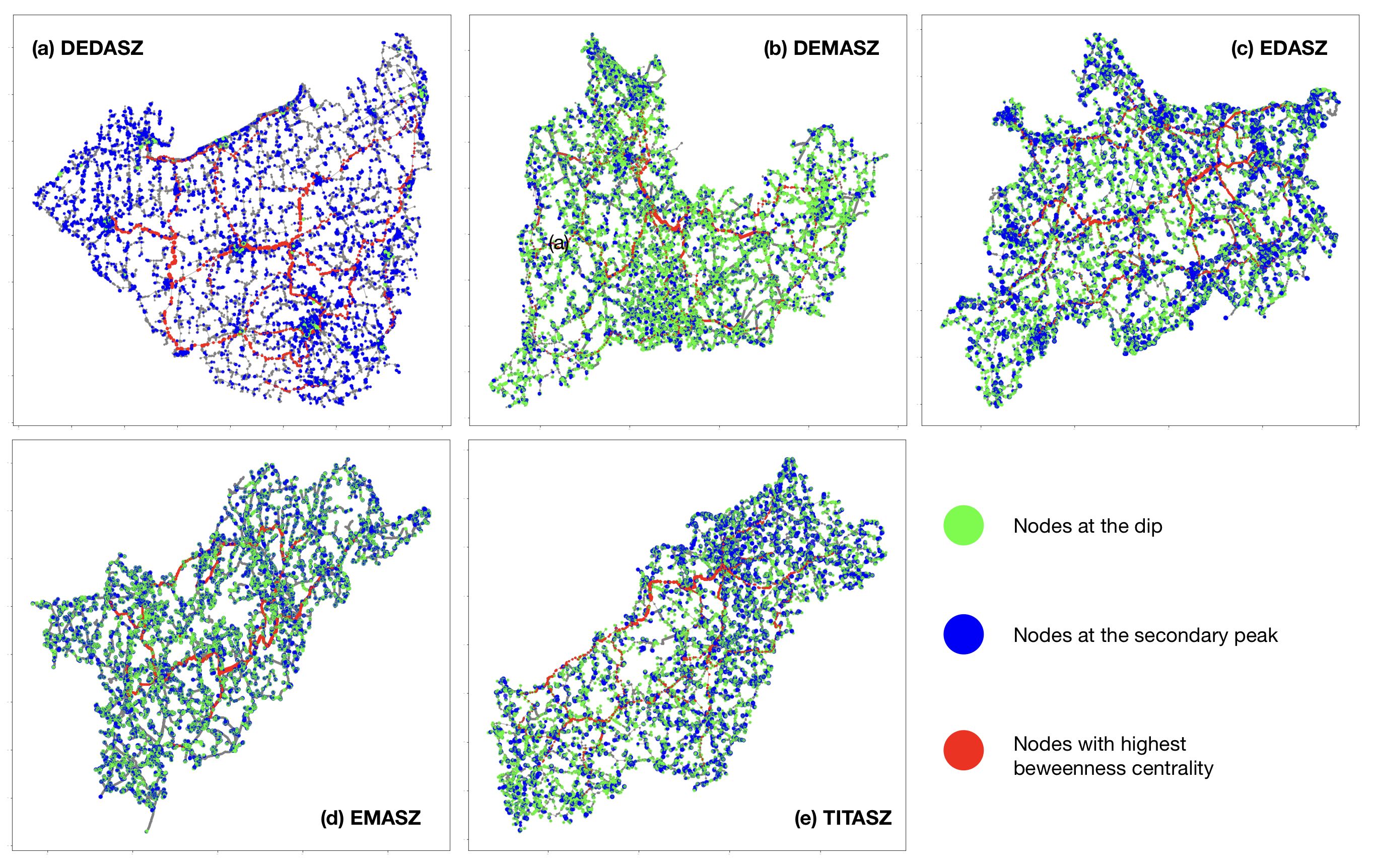}
\caption{\textit{Spatial distribution of nodes} found in the dip (green), peak (blue), and tails (red) of the BC distributions of the electric distribution networks found in Figure~\ref{fig: centralities}(c).}
\label{fig:centralities_maps}
\end{figure}

\begin{figure}[H]
\centering
\includegraphics[width = 0.9\linewidth]{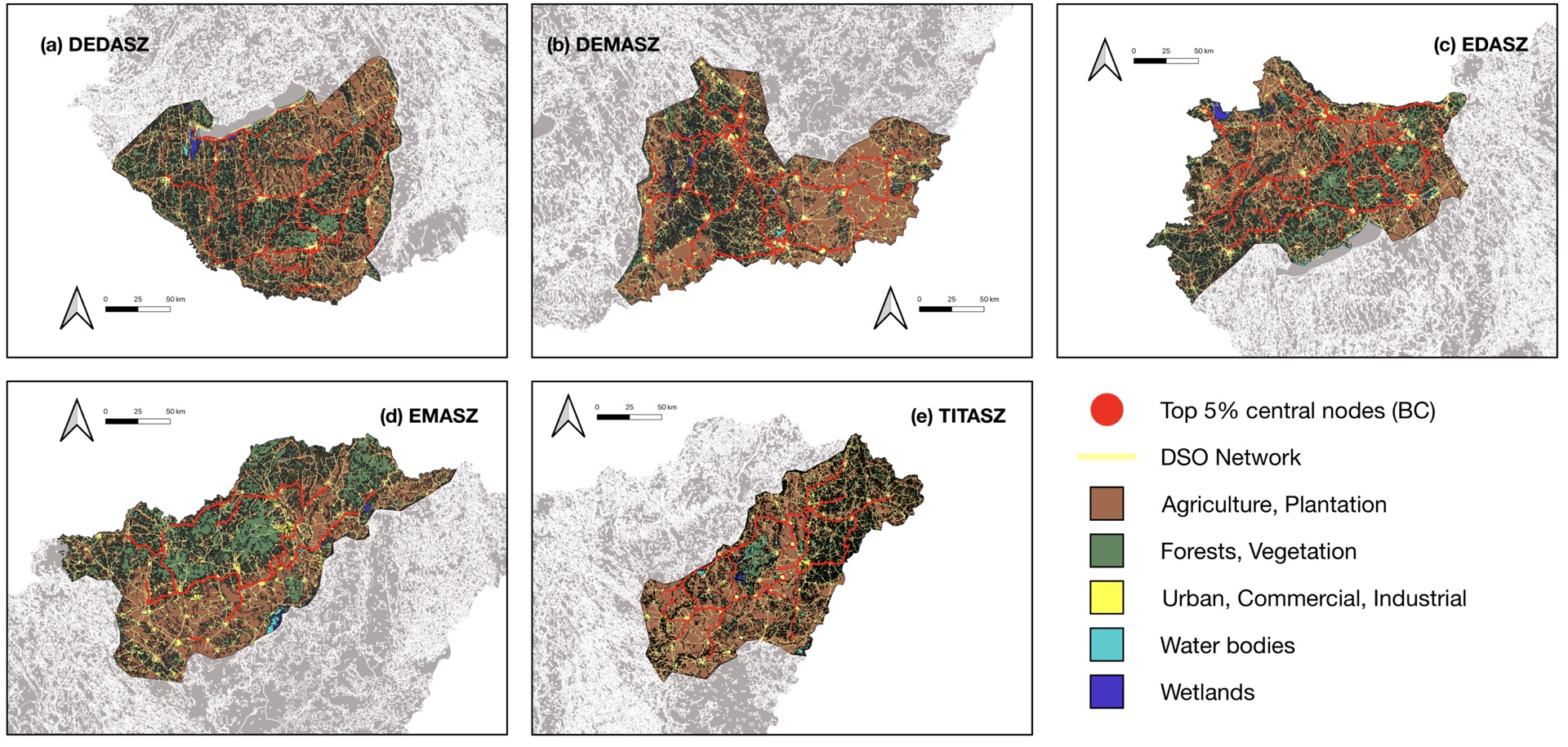}
\caption{\textit{Landcover information of regions catered by the DSO networks and the location of the most central nodes.}}
\label{fig:landcover_bc}
\end{figure}

In the context of powergrids, it is interesting to identify the location of the nodes that give rise to this observed universal behavior of \textit{"dip-second peak"}. In Figure~\ref{fig:centralities_maps}, we plotted the nodes with $BC$-values located at the dip, secondary peak, and tails of the distributions found in Figure~\ref{fig: centralities}(c), respectively. Here, we see that the nodes at the secondary peak, denoted by blue, are the centers of the clusters. These nodes are feeder substations that act as the hub that feeds multiple radial lines. In contrast, the nodes found at the dip, denoted by green, are the peripheral nodes of these clusters, which \st{may }represent the ends of the radial lines that serve local distribution transformers, which step down the voltage again for final delivery. These connect to residential, commercial, and small industrial consumers. Finally, the nodes found at the tails (largest betweenness values), denoted by red, make up a trail of nodes that frames and connects the sections of the regions being catered to by the Hungarian DSO networks. These can be major substations, transformers, or important transmission lines that connect different parts of the networks. True enough, if we look closely into Figure~\ref{fig:landcover_bc}, we may observe that this series of red nodes passes through and connects the urban sections of the regions. This makes them crucial in the connectivity and load flow; failure or overload at such a node can cause widespread disruption.

Overall, the network structure is well-balanced, with no extreme hubs and most nodes sharing similar roles in terms of connectivity and centrality. This structure contributes to efficient communication and resilience, as the network is not overly dependent on any single node for connectivity. While the presence of critical nodes with high-BC indicates some vulnerability, the truncation of their centrality limits the potential for catastrophic failures. In summary, the network is moderately resilient and efficient, with balanced connectivity and a few critical nodes that are important for maintaining flow, but not so central that their removal would lead to significant disruption. This is a result of the above-discussed cost-optimizing design approach~\cite{Kaiser2020} optional looping is provided with tie-lines,switches, among others to increase resilience of these networks,  which leads to a small proportion of bottleneck nodes and very few alternative routes.

\section{Conclusion and Recommendations}
\label{sec:conclusion}

Characterizing MV electric distribution networks is critical for power system planning, reliability, optimization, and resilience~\cite{BaranIEEEPOWERDELIVERY1989, AbeysingheAPPLIEDENERGY2018, HinesINTSYSSCI2010}. Ensuring these is important as MV networks serve as the backbone of electricity distribution, linking HV transmission systems to LV consumers. In this work, we provided a comprehensive spatial and topological characterization of five (5) Hungarian DSO networks, which have been repaired (i.e., corrections were made on vector data to ensure that network elements appeared as continuous lines)and thus have never been studied before.

Here, we found that the number of microvillages and population density are more influential on network structure and efficiency than total population, as indicated by the magnitude (regardless of sign) of the correlations, indicating that how people are distributed across a region matters more than how many people there are. For example, DEDASZ is observed to be highly fragmented and dispersed by having 842 settlements (91\% are microvillages) with low level of powerline density. However, its population density is the lowest. On the other hand, DEMASZ is found to have moderate population density and no microvillages, hinting at a more centralized state. Both are found to have a moderate performance in terms of electric supply reliability and quality.

Additionally, EMASZ is the most robust and efficient, combining reliability with quick recovery which is unexpected because its metrical and topological properties suggest that it is sparse by having high population density, low powerline density, and low average degree, clustering coefficient, and modularity, which all points to or suggestive of challenges in network coverage. On the other hand, TITASZ and EDASZ are considered to exhibit a balance in size and connectivity. Ironically, TITASZ, despite fast restoration, struggles with system reliability due to frequent faults and longer outages. Finally, EDASZ exhibits the worst performance across most of the electric supply quality metrics.

Despite notable differences in geographic layout and consumer distribution, we identify statistically consistent patterns across several key metrics, including degree, BC, and powerline length distributions. These findings confirm the influence of common underlying design principles or optimization constraints, potentially indicating universal structural tendencies in MV network design. The results provide insight into the organization of real-world distribution systems and offer a basis for improved planning, risk mitigation, and system optimization in future grid developments.

\section*{Acknowledgements}

B\'alint Hartmann acknowledges the support of the
Bolyai János Research Scholarship of the Hungarian
Academy of Sciences (BO/131/23).

\section*{Author contributions statement}

\textbf{M.T.C:} Writing – Original draft, Formal analysis, Data curation, Visualization \textbf{B.H.:} Writing – original draft, Validation, Supervision, Resources, Project administration, Methodology, Investigation, Funding acquisition, Formal analysis, Conceptualization. \textbf{T.E.:} Writing – original draft, Data curation. \textbf{T.S.:} Writing – original draft, Visualization, Validation.  All authors reviewed the manuscript.
\end{document}